\documentclass[aps,superscriptaddress,,twocolumn,showpacs]{revtex4-1}
\pdfoutput=1
\usepackage{color}
\usepackage{amsmath, amsthm, amsfonts}    
\usepackage{amssymb}
\usepackage{mathptmx} 
\usepackage{dsfont}
\usepackage{cancel}
\usepackage{tikz}
\usepackage[unicode=true,pdfusetitle,
 bookmarks=true,bookmarksnumbered=false,bookmarksopen=false,
 breaklinks=true,pdfborder={0 0 0},backref=false,colorlinks=true,citecolor=blue]{hyperref}
\usepackage{graphicx}   
\usepackage{epstopdf}
\usepackage[caption=false]{subfig}       
\usepackage{bm}            
\usepackage[normalem]{ulem}  

\newlength\imagewidth
\newlength\imagescale
\def\be{\begin{eqnarray}}
\def\ee{\end{eqnarray}}

\def\E{{\bf E}}
\def\H{{\bf H}}

\def\im{{\rm i}}

\definecolor{JOT-color}{named}{blue}
\definecolor{CSF-color}{named}{orange}
\definecolor{DRA-color}{named}{magenta}
\definecolor{AGE-color}{named}{brown}
\definecolor{JAG-color}{named}{red}
\definecolor{GMT-color}{named}{purple}
\definecolor{NDS-color}{named}{teal}

\begin{document}

\title{Kerker Conditions Upon Lossless, Absorption, and Optical Gain Regimes}

\author{Jorge Olmos-Trigo}
\email{jolmostrigo@gmail.com}
\affiliation{Donostia International Physics Center (DIPC),  20018 Donostia-San Sebasti\'{a}n, Basque Country, Spain}

\author{Cristina Sanz-Fern\'andez}
\affiliation{Centro de F\'{i}sica de Materiales (CFM-MPC), Centro Mixto CSIC-UPV/EHU, 20018 Donostia-San Sebasti\'{a}n, Spain}

\author{Diego R. Abujetas}
\affiliation{Donostia International Physics Center (DIPC),  20018 Donostia-San Sebasti\'{a}n, Basque Country, Spain}
\affiliation{Instituto de Estructura de la Materia (IEM-CSIC), Consejo Superior de Investigaciones Cient\'{\i}ficas, Serrano 121, 28006 Madrid, Spain}

\author{Jon Lasa-Alonso}
\affiliation{Donostia International Physics Center (DIPC),  20018 Donostia-San Sebasti\'{a}n, Basque Country, Spain}
\affiliation{Centro de F\'{i}sica de Materiales (CFM-MPC), Centro Mixto CSIC-UPV/EHU,  20018 Donostia-San Sebasti\'{a}n,  Spain}

\author{Nuno de Sousa}
\affiliation{Donostia International Physics Center (DIPC),  20018 Donostia-San Sebasti\'{a}n, Basque Country, Spain}

\author{Aitzol Garc\'{i}a-Etxarri}
\affiliation{Donostia International Physics Center (DIPC),  20018 Donostia-San Sebasti\'{a}n, Basque Country, Spain}
\affiliation{Centro de F\'{i}sica de Materiales (CFM-MPC), Centro Mixto CSIC-UPV/EHU,  20018 Donostia-San Sebasti\'{a}n,  Spain}

\author{Jos\'e A. S\'anchez-Gil}
\affiliation{Instituto de Estructura de la Materia (IEM-CSIC), Consejo Superior de Investigaciones Cient\'{\i}ficas, Serrano 121, 28006 Madrid, Spain}

\author{Gabriel Molina-Terriza}
\affiliation{Donostia International Physics Center (DIPC),  20018 Donostia-San Sebasti\'{a}n, Basque Country, Spain}
\affiliation{Centro de F\'{i}sica de Materiales (CFM-MPC), Centro Mixto CSIC-UPV/EHU,  20018 Donostia-San Sebasti\'{a}n,  Spain}
\affiliation{IKERBASQUE, Basque Foundation for Science, 48013 Bilbao, Spain}

\author{Juan Jos\'e S\'aenz}
\affiliation{Donostia International Physics Center (DIPC),  20018 Donostia-San Sebasti\'{a}n, Basque Country, Spain}
\affiliation{IKERBASQUE, Basque Foundation for Science, 48013 Bilbao,  Spain}

\begin{abstract}
The directionality and polarization of light show peculiar properties when the scattering by a dielectric sphere can be  described exclusively by electric and magnetic dipolar modes. Particularly, when these modes oscillate \emph{in-phase} with equal amplitude, at the so-called first Kerker condition, the zero optical backscattering condition emerges for non-dissipating spheres. However, the role of absorption and optical gain in the first Kerker condition remains unexplored. 
In this work, we demonstrate that either absorption or optical gain precludes  the first Kerker condition and, hence, the absence of backscattered radiation light, regardless of the particle's size, incident wavelength, and incoming polarization.
Finally, we derive the {necessary} prerequisites of the second Kerker condition of the zero forward light scattering, 
finding that optical gain is {a compulsory requirement}. 
\end{abstract}

\maketitle
{In 1983, Kerker, Wang, and Giles predicted that, under plane wave illumination, magnetic spheres with equal relative  permittivity $\epsilon$ and  permeability $\mu$ radiate no light in the backscattering direction \citep{kerker1983electromagnetic}. 
They also concluded that if $\epsilon = (4 - \mu) / (2 \mu +1)$ for nano-spheres, this zero optical light scattering condition happened at the forward direction.}

Three decades later, a renewed version of these ideas was proposed for subwavelength dielectric spheres ($\mu=1$) of high refractive index (HRI) materials~\cite{nieto2011angle}, reinvigorating the interest on  these light scattering conditions. Notably, the scattering properties of these HRI nano-spheres can be fully described  by dipolar modes via the first electric and magnetic Mie coefficients, without a spectral overlap from higher-order modes for certain ranges of the electromagnetic spectrum~\cite{garcia2011strong,kuznetsov2016optically}. 
In terms of the electric and magnetic  scattering phase-shifts~\cite{hulst1957light}, these coefficients generally read as,
\begin{align}\label{Mie_dip}
a_l = \im \sin \alpha_l e^{-\im\alpha_{l}}, && b_l = \im \sin \beta_l e^{-\im \beta_{l}},
\end{align}
respectively, where $\alpha_l$ and $\beta_l$ are real in the absence of losses or optical gain. At the \emph{first Kerker condition}~\cite{nieto2011angle}, given by $\alpha_1 = \beta_1 \Longleftrightarrow a_1 =b_1$, the electric and magnetic dipolar modes oscillate \emph{in-phase} with equal amplitude. This  optical response drives to destructive interference between the scattered fields at the backscattering direction, which is commonly referred  to as \emph{zero optical backscattering condition}~\cite{kerker1983electromagnetic}. This anomalous light scattering condition was first experimentally measured in the limit of small particle in the microwave regime for ceramic spheres~\cite{geffrin2012magnetic} and, soon after, in the visible spectral range for HRI Si~\cite{fu2013directional} and GaAs nano-spheres~\cite{person2013demonstration}. However, recent results suggest that the concept of small particle is sufficient, but not necessary, to guarantee a dipolar response in the optical scattering of an object~\cite{olmos2020unveiling}. Consequently, these aforementioned backscattering anomalies could also be measured on larger dielectric particles. Interestingly,  the absence of backscattered light emerges at the first Kerker condition for dipolar particles regardless of the incoming polarization~\cite{luk2015optimum, alaee2015generalized, yang2020electromagnetic}. 
However, for incoming beams with well-defined helicity (handedness of the fields)~\cite{calkin1965invariance,olmos2019sectoral, wei2020momentum}, the absence of backscattered light arises for cylindrical symmetrical particles when the EM helicity is a preserved quantity after scattering~\cite{zambrana2013duality,fernandez2013electromagnetic}. Conservation of helicity has proven crucial in many applications such as enhanced chiral light-matter interactions~\cite{garcia2013surface, eismann2018exciting, alpeggiani2018electromagnetic, poulikakos2019optical, garcia2019enhanced, feis2020helicity, graf2019achiral, lasa2020surface}, or in the spin-orbit interactions of light \cite{schwartz2006conservation,tkachenko2014helicity, bliokh2015spin, rafayelyan2016reflective, nechayev2019orbital, abujetas2020spin}.
In this vein, it has been reported that from a relatively simple far-field measurement of the EM helicity at a right angle, the radiation pattern of the dipolar particle is inferable~\cite{olmos2019asymmetry}. 
This phenomenon arises since the asymmetry parameter ($g$), which encodes the particle's optical response, is equivalent to the EM helicity at the direction perpendicular to the incoming wave when the object is excited by a beam with well-defined helicity ($\sigma = \pm 1$), namely, $\Lambda_{\pi/2} = 2 \sigma g$~\cite{olmos2019asymmetry}. This relation straightforwardly links the EM helicity with the $g$-parameter, which appears in multiple branches of physics such as optical forces~\cite{nieto2010optical,gomez2012electric, xu2020kerker}, light transport phenomena~\cite{gomez2012negative,naraghi2015directional, varytis2020negative}  or wavelength-scale errors in optical localization~\cite{araneda2018wavelength}. Remarkably, this wavelength's error limit can be drastically surpassed at the first Kerker condition for dipolar particles, where an optical vortex arises in the backscattering region~\cite{olmos2019enhanced}.

In contrast to the first Kerker condition, the zero optical scattering condition in the forward direction, given for dipolar particles by $a_1 = -b_1$, is precluded by the optical theorem for lossless spheres~\cite{kerker1978electromagnetic,alu2010does,liu2018generalized}.
As an alternative, the generalized second Kerker condition (GSKC), mathematically expressed for dipolar particles as $\alpha_1 = -\beta_1$, was proposed as an approximated condition
for non-dissipating spheres that might guarantee the maximum backward/forward scattering ratio while respecting energy conservation~\cite{nieto2011angle}. Indeed, the GSKC is the optimal backward light scattering condition; however, it does not generally imply a nearly-zero optical forward scattering~\cite{olmos2020optimal}, contrary to what could be expected from previous works~\cite{nieto2011angle, pors2015unidirectional,decker2016resonant, gao2017optical}.

Most research on this topic is dedicated to the optical response of lossless HRI nano-spheres in the dipolar regime. However, the role of absorption and optical gain remains unexplored in the context of Kerker conditions and its abovementioned anomalous light scattering conditions~\cite{olmos2019role}. 
In this work, we  demonstrate analytically that either losses or optical gain inhibit the first Kerker condition for dielectric  Mie spheres regardless of the  particle's size, incident wavelength, incoming polarization, and multipole order. Consequently, these results unveil a hidden  connection between energy conservation, mathematically expressed in terms of the optical theorem, and the first Kerker condition.  As a result, we show that the EM helicity cannot be preserved after scattering by an arbitrary dielectric sphere in the presence of losses or optical gain. Hence, neither can the zero optical backscattering condition be fulfilled in that scenario.  In particular, for a Germanium (Ge) sphere in the  dipolar regime, we quantify the gradual drift from the ideal zero optical backscattering condition as the absorption rate is increased.
Finally, we prove that optical gain is mandatory to reach the zero forward light scattering condition. 

Mie theory~\cite{hulst1957light}  {gives an} exact solution of
Maxwell's equations for  a
spherical particle in a homogeneous medium under plane wave  illumination. It allows {writing} the extinction, scattering, and absorbing efficiencies
of the particle as
\begin{equation}\label{extincion}
Q_{\rm{ext}} = \frac{2}{x^2} \sum_{l= 1}^{\infty} \left( 2l +1 \right) \Re \{a_l +b_l \} = \sum_{l= 1}^{\infty} \left(Q^{a_l}_{\rm{ext}} + Q^{b_l}_{\rm{ext}} \right),
\end{equation}
\begin{equation}\label{scattering}
Q_{\rm{sca}} = \frac{2}{x^2} \sum_{l= 1}^{\infty} \left( 2l +1 \right) \left(  |a_l|^2 + |b_l|^2  \right) = \sum_{l= 1}^{\infty} \left(Q^{a_l}_{\rm{sca}} + Q^{b_l}_{\rm{sca}} \right),
\end{equation}
where $Q_{\rm{abs}} = Q_{\rm{ext}}  - Q_{\rm{sca}}$.

The efficiencies are dimensionless magnitudes  given by the ratio between the cross section  and the geometrical area, $Q = \sigma / \pi R ^2 $, where $R$ is the radius of the particle.
Here, $x = kR$ is the size parameter, $k$ = ${\rm{m}}_{\rm{h}} k_0 = {\rm{m}}_{\rm{h}} \left(2 \pi \right) / \lambda_0$, being $\lambda_0$ the wavelength in vacuum and   ${\rm{m}}_{\rm{h}}$ the  refractive index of the external medium. The Mie coefficients, $a_l$ and $b_l$, can be expressed in terms of the scattering phase-shifts (see Eq.~\eqref{Mie_dip}) by~\cite{hulst1957light}, 
\begin{equation}\label{etan}
\tan \alpha_l = -\frac{S'_l({\rm{m}}x)S_l(x) -{\rm{m}} S_l({\rm{m}}x)S'_l(x)}{S'_l({\rm{m}}x)C_l(x) -{\rm{m}} S_l({\rm{m}}x)C'_l(x)},
\end{equation}
\begin{equation}\label{mtan}
\tan \beta_l = -\frac{{\rm{m}}S'_l({\rm{m}}x)S_l(x) -S_l({\rm{m}}x)S'_l(x)}{{\rm{m}}S'_l({\rm{m}}x)C_l(x)-  S_l({\rm{m}}x)C'_l(x)}.
\end{equation}
Here ${\rm{m}} = {\rm{m}_p} / {\rm{m}}_h$ is the refractive index contrast, where ${\rm{m}_p}$ is the refractive index of the particle while $S_l(z) = \sqrt{\frac{\pi z }{2}} J_{l+\frac{1}{2}}(z)$ and $C_l(z) = \sqrt{\frac{\pi z }{2}} N_{l+\frac{1}{2}}(z)$ are the Riccati-Bessel functions, 
where $J_{l+\frac{1}{2}}(z)$  and $N_{l+\frac{1}{2}}(z)$ are the Bessel and Neumann functions, respectively. 

According to Eqs.~\eqref{etan} and \eqref{mtan},  the first Kerker condition, in which the electric and magnetic dipolar modes oscillate \emph{in-phase} with identical amplitude,  can be obtained either when $S_1({\rm{m}}x) = 0$ (nodes of first kind) or when $S'_1({\rm{m}}x) = 0$ (nodes of second kind) \citep{hulst1957light}. However, for  complex values of the refractive  index contrast~\footnote{Note that the argument ${\rm{m}} x = {\rm{m}}_p k_0 R$ does not depend on the refractive index of the external medium.}, i.e.,  $\Im \{ {\rm{m}}\} \neq 0$, which corresponds either with absorption ($\Im \{ {\rm{m}} \}>0$) or active media ($\Im \{ {\rm{m}} \}<0$), these nodes are unreachable. We will prove this, and generalize it for arbitrary multipolar modes, using the following Lemmas:
\begin{enumerate}  
\item \label{L1} When $v > -1$  the zeros of $J_v(z)$ are all real~\cite{watson1995treatise},
\end{enumerate}
\begin{enumerate} 
\setcounter{enumi}{1}
\item \label{L2} When $v > -1$ and $a,b \in \mathbb{R} $, then $a J_v(z) + b z J'_v(z)$ has all its zeros real, except
 when $a/b + v <0$~\cite{watson1995treatise}.
\end{enumerate}

Lemma~\ref{L1} directly implies that the node of the first kind, $S_l({\rm{m}} x) = 0$, are inhibited for spheres with either gain or loss since the zeros of the Bessel functions occur for exclusively real arguments, while in these cases $\Im(m) \neq 0$ . On the other hand,  the node of second kind, given by 
\begin{equation} \label{Second} S'_l({\rm{m}}x) = J_{l+\frac{1}{2}}({\rm{m}}x) + 2 {\rm{m}}x \  J'_{l+\frac{1}{2}}({\rm{m}}x) = 0, 
\end{equation}
cannot be satisfied for $\Im \{ {\rm{m}} \} \neq 0$, since following Lemma~\ref{L2} with $a=1$, $b=2$, and $v = l+ 1/2$ the condition $a/b + v <0$ is inaccessible because $l \geq 1$ ~\cite{hulst1957light}.

The immediate physical consequence of these Lemmas is straightforward: either absorption or optical gain inhibits the emergence of the first Kerker condition. It is important to {note} that the validity of these conclusions holds regardless of the  particle size, incident wavelength, (complex) refractive index contrast, and polarization of the incoming light. Remarkably, this result is valid for any multipole order $l$. In short, we can conclude that $a_l \neq b_l \; \forall \: l$ when $\Im \{ {\rm{m}} \} \neq 0$, making this demonstration general. 

Interestingly, these conclusions can also be understood by analysing the extinction and scattering efficiencies  arising from electric and magnetic modes. In the presence of losses or gain, the extinction and scattering efficiencies of an arbitrary electric multipole $l$  cannot be identical to the magnetic counterpart of the same multipole $l$. According to the right side of Eqs.~\eqref{extincion} and \eqref{scattering}, this phenomenon implies the following: Taking into account that in the presence of gain or losses $Q_{abs} \neq 0$, if $Q^{a_l}_{\rm{sca}} = Q^{{b_l}}_{\rm{sca}}$ then $Q^{a_l}_{\rm{ext}} \neq Q^{{{b}}_l}_{\rm{ext}}$. {These relations imply that if $\Im \{ m \} \neq 0$ the electric and magnetic modes cannot {simultaneously}  oscillate in-phase  with equal amplitude, unveiling a connection between the first Kerker condition and energy conservation.}

\begin{figure}[t!]
\includegraphics[width=1 \columnwidth]{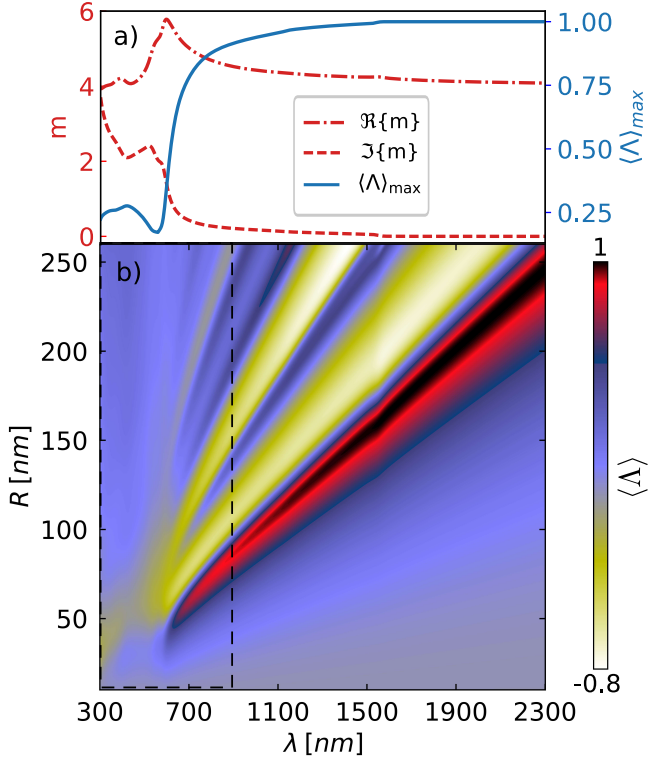}
\captionsetup{justification= raggedright}
\caption{(a) Real (dash-dotted red) and imaginary part (dashed-red) of the refractive index contrast (${\rm{m}}$) vs the incident wavelength ($\lambda$) for a Ge sphere. Maximum of the expected value of the EM helicity in solid-blue, $\langle \Lambda \rangle_{max}$, for a Ge sphere vs $\lambda$ under plane wave illumination with $\sigma =+1$. (b) Color map of  $\langle \Lambda \rangle$ vs $\lambda$ and particle's size ($R$) for a Ge sphere under plane wave illumination with $\sigma = +1$. The visible range is encompassed by a dashed rectangle. As mentioned in the text, in this region helicity conservation is never fulfilled. }
\label{Ge_color}
\end{figure}

To get a deeper insight into these results, let us calculate the expected value of EM helicity after scattering.  
The scattered fields outside the sphere, decomposed in components of well-defined EM helicity~\cite{aiello2015note}, i.e., $\E_{\text{sca}} = \E_{\text{sca}}^+ + \E_{\text{sca}}^-$ with $\bm{\Lambda}\E^{\sigma}_{\text{sca}} = \sigma \E^{\sigma}_{\text{sca}} $, can be written in terms of ``outgoing'' vector spherical wave functions, $\bm{\Phi}_{lm}^{\sigma'}$ (defined in \cite{olmos2019sectoral}) as
\be \label{sca_field}
\E_{\text{sca}}^\sigma &=& E_0 \sum_{l=1}^{\infty} \sum_{m=-l}^{+l}  D_{lm}^{ \sigma} \boldsymbol{\Phi}_{lm}^{\sigma},  \\
\begin{pmatrix} D_{lm}^+ \\ D_{lm}^- \end{pmatrix} &=&  
-\begin{pmatrix}\label{sca_coef} [a_l+b_l] & [a_l-b_l] \\ [a_l-b_l] & [a_l +b_l]   \end{pmatrix} 
\begin{pmatrix} C_{lm}^+ \\   C_{lm}^- \end{pmatrix}, \\
Z\H_{\text{sca}}^\sigma &=& -\im {\bm{\Lambda}} \E_{\text{sca}}^\sigma ,
 \label{magnetic}
\label{multipolar_scatter} 
\ee
where $C^{\sigma}_{lm}$ are the expansion coefficients  of  the  incoming wave in a basis of vector spherical harmonics.

Under illumination by a circularly polarized plane wave with well-defined helicity ($\sigma = \pm 1$) and AM  in the wave's propagation direction  $J_z = m = \sigma$~\cite{olmos2019sectoral}, it can be shown that the expected value of the EM helicity is given by~\cite{olmos2020unveiling}
\begin{align} \label{hel_mean} 
\langle \Lambda \rangle = \frac{\int \E^*
_{\text{sca}} \cdot {\bm{\Lambda}} \E_{\text{sca}} \; d \Omega}{\int \E^*
_{\text{sca}} \cdot  \E_{\text{sca}} \; d \Omega} = 2 \sigma \left[ \frac{\sum_{l=1}^\infty  \left(2l+1 \right)   \Re \{a_l b^*_l \} }{\sum_{l=1}^\infty \left(2l+1 \right) \left(  |a_l|^2 + |b_l|^2 \right) } \right].
\end{align}
From Eq.~\eqref{hel_mean} it is straightforward to notice that {in the presence of gain or losses, since $a_l \neq b_l \; \forall \: l$},  the EM helicity is not preserved, namely, $|\langle \Lambda \rangle| \neq 1$.

Figure~\ref{Ge_color} summarizes quantitatively this conclusion for a Ge sphere of different radii. { For $\Im \{{\rm{m}} \} >0$, corresponding to the visible spectral range (see Fig.~\ref{Ge_color}a)), the EM helicity is far from being preserved, regardless of the size of the Ge sphere in the entire visible spectral range which corresponds to the dashed rectangle in Fig.~\ref{Ge_color}b). This phenomenon can be inferred from $\langle \Lambda \rangle_{\rm{max}}$, which is obtained, by finding for each of the incident wavelengths, the radius that maximizes the EM helicity.}  Contrary, in the telecom spectral range, where losses are negligible (see Fig.~\ref{Ge_color}a)), the maximum value of the EM helicity is preserved at the first Kerker condition, $\langle \Lambda \rangle_{\rm{max}} \approx 1$.

\begin{figure}[t!]
\includegraphics[width=1\columnwidth]{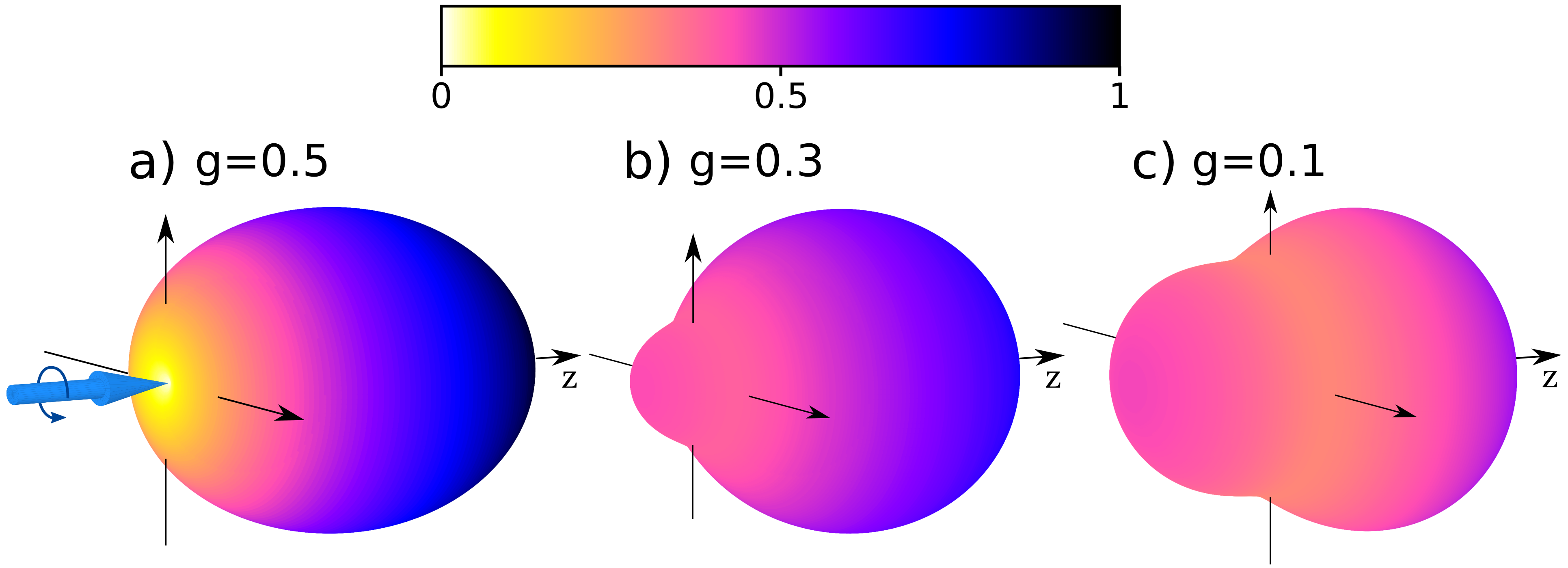}
\captionsetup{justification= raggedright}
\caption{Normalized scattering  patterns by Ge spheres calculated from Eq.~\eqref{dcross} in the telecom spectral regime (a) [$\lambda = 2100$ nm  and $R=223$ nm] and visible spectral range (b) [$\lambda = 632$ nm  and $R=48$ nm] and (c) [$\lambda = 575$ nm  and $R=35$ nm]. The $g$-parameter is given by $g=0.5, 0.3, 0.1$, respectively, in the dipolar regime.  }
\label{Loss}
\end{figure}

To get insights into the relevance of the breaking of the first Kerker condition due to absorption effects in the scattering radiation pattern, let us now consider the differential scattering cross section~\cite{hulst1957light},
\begin{equation}
\frac{d \sigma_{\rm{s}}}{d \Omega} =  \lim_{kr \rightarrow \infty} r^2 \frac{{\bf{S} \cdot \hat{\bf{r}}}}{S_0}.
\end{equation}
Here ${\bf{S} = \Re \{ \E \times \H \}} / 2$ denotes the scattered Poynting vector, $S_0$ refers to the amplitude of the incoming Poynting vector amplitude and $\bf{\hat{r}}$ is the radial unit vector. By taking into account  Eqs.~\eqref{sca_field}--\eqref{magnetic}, when just retaining the dipolar contribution ($l=1$), it can be shown that the (integral-normalized) differential scattering cross section reads as
\begin{align} \label{dcross}
\frac{d \sigma_{\rm{s}}}{d \Omega}  = \frac{3}{8 \pi} \left( \frac{1 + \cos^2 \theta}{2} + 2g\cos \theta \right),
\end{align}
where $g$ is the asymmetry parameter in the dipolar regime~\cite{gomez2012negative}.

From  Eq.~\eqref{dcross} it is noticeable that at the first Kerker condition, when the EM helicity is preserved (see Eq.~\eqref{hel_mean} for $l=1$), the asymmetry parameter is maximized in the dipolar regime, i.e., $g =0.5$. In this scenario, it can be inferred from  Eq.~\eqref{dcross} that there is no  radiation in the backscattering direction ($\theta = \pi$)~\cite{person2013demonstration,geffrin2012magnetic, fu2013directional}. However, as it was previously deduced in the presence of losses or optical gain,  the zero optical backscattering condition cannot emerge as a result of the breaking of the first Kerker condition that imposes both $g<0.5$ and $|\Lambda|<1$.
As an illustrative example, we show in Fig.~\ref{Loss} the gradual loss of the zero optical backscattering condition (see Fig.~\ref{Loss}a)), as the absorption rate is increased for a Ge sphere (see Fig.~\ref{Loss}b) and Fig.~\ref{Loss}c)). As can be deduced, it is easy to see how the zero optical backscattering condition is lost in the case of lossy spheres. 

\begin{figure}[t!]
\includegraphics[width=1 \columnwidth]{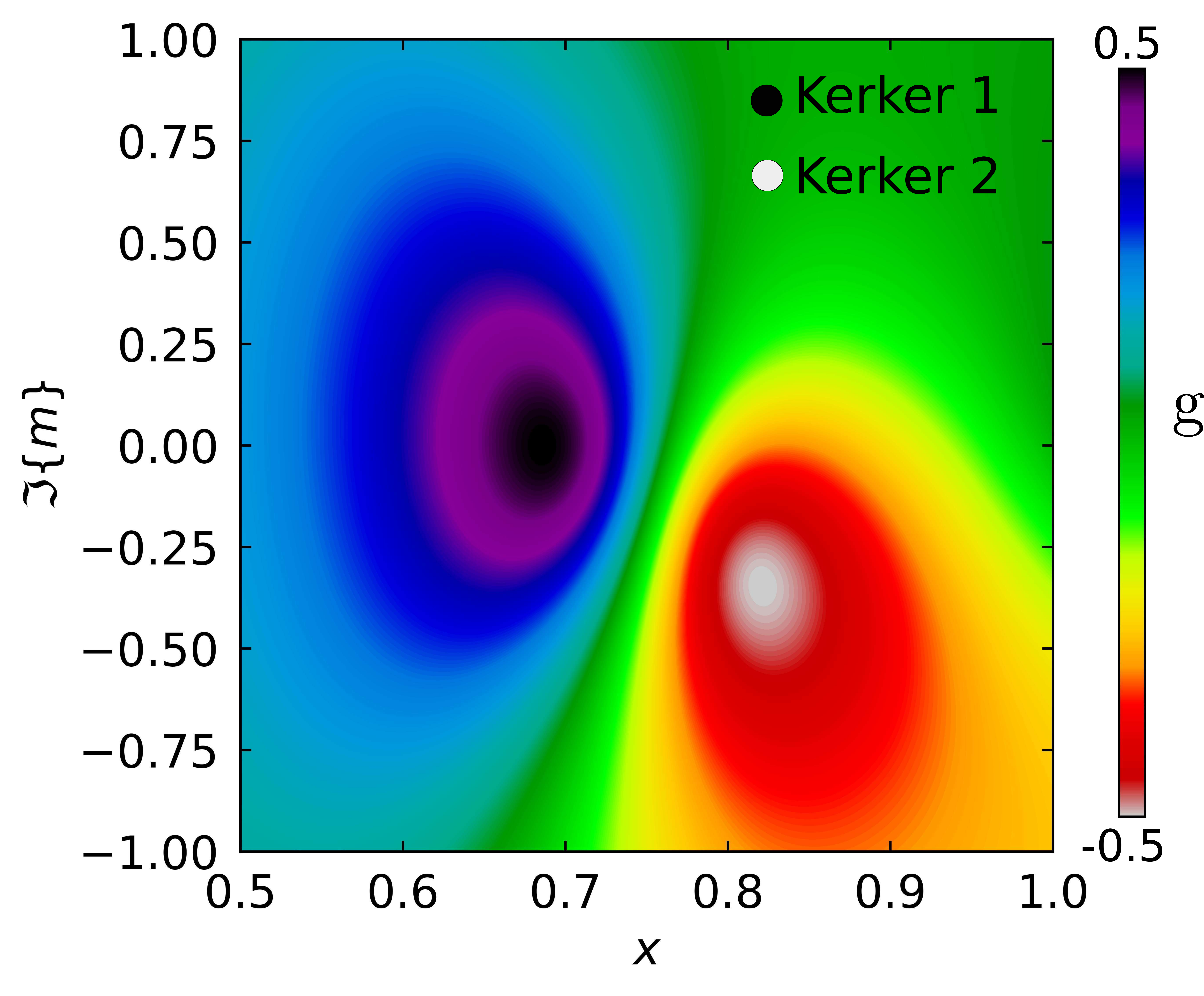}
\captionsetup{justification= raggedright}
\caption{$g$-parameter vs the imaginary part of the refractive index contrast, $\Im \{ {\rm{m}} \}$, and the $x=ka$ size parameter under well-defined EM helicity ($\sigma =+1$) plane wave illumination. In this range, the optical response is purely of dipolar nature. The first and second Kerker conditions are depicted by black and white circles, respectively. }
\label{Ge_yingyang}
\end{figure}

Finally, let us briefly analyse the second Kerker condition,
given by $a_1 =-b_1$. Let us recall that according to Eq.~\eqref{hel_mean} in the dipolar limit, the EM helicity flips its value from $\langle \Lambda \rangle = +\sigma$ to $\langle \Lambda \rangle = -\sigma$, and the $g$-parameter is minimized, $g = -0.5$, leading to zero optical light scattering in the forward direction (see  Eq.~\eqref{dcross}). 
According to  Eq.~\eqref{Mie_dip}, the second Kerker condition implies both
\begin{align} \label{Kerker2}
\sin2\alpha_1 = -\sin2 \beta_1  && \text{and} && \sin^2 \alpha_1 = -\sin^2 \beta_1.
\end{align}
It is straightforward to notice that lossless spheres, where $\Im \{ {\rm{m}}\} = 0$, cannot satisfy the second Kerker condition since in that scenario $\alpha_l, \beta_l \in \mathbb{R}$ and then, the right side of Eq.~\eqref{Kerker2} is  unreachable. 
{Interestingly, the second Kerker condition leads to the anti-duality condition for dipolar particles, phenomenon that cannot be achieved for non-active media~\cite{zambrana2013dual, gutsche2018optical}, as we have demonstrated.}

Figure~\ref{Ge_yingyang} illustrates the $g$-parameter for a Ge-like sphere ($m=4$) versus the size parameter, $x =kR$, and the imaginary part of the  refractive index contrast, $\Im \{{\rm{m}} \}$,  under plane wave illumination with well-defined helicity ($\sigma =+1$). In this regime, the optical response is almost entirely dipolar and, as a result, the asymmetry parameter is in essence the same magnitude as the expected value of the EM helicity (and $\Lambda_{\pi/2}$), $\langle \Lambda \rangle = 2g$~\cite{olmos2019asymmetry}.  As previously mentioned, the first Kerker condition ($a_1 =b_1$) arises in a lossless regime ($\Im \{{\rm{m}} \} = 0$) at $x \sim 0.675$. This specific size parameter corresponds to the first Kerker condition appearing in  Fig.~\ref{Loss}a). As expected, the first Kerker condition does not emerge for $\Im \{{\rm{m}} \} \neq 0$.  On the other hand, the second Kerker condition does not appear for $\Im \{{\rm{m}} \} = 0$, in agreement with Eq.~\eqref{Kerker2}. In fact, it arises if and only if optical gain is being pumped onto the system. In the particular case of the Ge-like sphere, it emerges for $\Im \{{\rm{m}} \} \sim -0.3$ and $x \sim 0.825$, as can be reckoned from  Fig.~\ref{Ge_yingyang}.

In conclusion, we have rigorously demonstrated that either losses or optical gain inhibit the appearance of the first Kerker condition for dielectric spheres. Therefore, in lossy systems, the phenomena associated with the first Kerker condition are greatly modified, and one should carefully analyze the situations where the magnetic and electric extinctions cross to be able to derive conclusions. As a direct consequence of our analysis, we have shown that the EM duality restoration, identified  through the conservation of the EM helicity and, hence, a null optical backscattering condition, cannot be achieved in the presence of losses. In a spin-orbit framework, this phenomenon precludes the full  angular momentum exchange from spin to orbit after scattering, inhibiting the emergence of an optical vortex in the backscattering direction. Furthermore, we have studied the gradual loss of the zero optical backscattering condition for a Ge sphere as the absorption is increased. We have also determined the conditions under which the second Kerker condition emerges and, therefore, the  zero forward optical scattering condition is met. The abovementioned statements can be summarized as follows: for the  imaginary part of the contrast index $\Im \{{\rm{m}} \} \neq 0$,  while the second Kerker condition is achievable, the first Kerker condition is inhibited. In this scenario, the zero optical forward light scattering can be achieved in the presence of optical gain. In contrast, for  $\Im \{{\rm{m}} \} = 0$, the first Kerker condition is obtainable while the second Kerker condition is unreachable. In this case, only the zero optical backscattering condition is reachable. 
Our analysis unveils an intriguing connection between the Kerker conditions and the energy conservation from fundamental principles, opening new insights into the so-called Mie theory.

The authors dedicate this work to the memory of their beloved colleague and friend, Prof. Juan Jos\'e S\'aenz, who passed away on March 22, 2020.

This research was supported by the Basque Government
(Project PI-2016-1-0041 and PhD Fellowship PRE-2018-
2-0252) and by the Spanish MINECO and MICINN and
European Regional Development Fund (ERDF) Projects:
 FIS2017-
91413-EXP, FIS2017-82804-P, FIS2017-87363-P, PGC2018-095777-B-C21 and
PhD Fellowship FPU15/ 03566.

\bibliography{New_era_18_06_2019}
\end{document}